\documentclass[aps,amsfonts,amsmath,prd,preprint,nofootinbib]{revtex4}
\newcommand{\beq}{\begin{equation}}
\newcommand{\eeq}{\end{equation}}

\usepackage{epsfig,bbm,cancel}
\usepackage[breaklinks=true]{hyperref}

\begin{document}

\title{Flux compactifications in Einstein--Born--Infeld theories}

\author{Handhika S. Ramadhan}
\email{hramad@ui.ac.id}
\author{Brian A. Cahyo}
\email{brian.agung@sci.ui.ac.id}
\author{Muhammad Iqbal}
\email{muhammad.iqbal06@sci.ui.ac.id}
\affiliation{Departemen Fisika, FMIPA, Universitas Indonesia, Depok 16424, Indonesia}

\def\changenote#1{\footnote{\bf #1}}

\begin{abstract}
We investigate the flux compactification mechanism in simple toy models of Einstein--Born--Infeld theories. These are the direct generalizations of the Einstein--Maxwell flux compactifications that recently gained fame as a toy model for tunneling in the landscape. Our investigation reveals that the Born--Infeld form does not significantly modify the qualitative result of the Einstein--Maxwell theory. For the case of Einstein--Higgs theory, however, we found that the effect of Born--Infeld nonlinearity is to render all $q>1$ extradimensional compactification unstable against semiclassical tunneling to nothing.  

\end{abstract}

\maketitle
\thispagestyle{empty}
\setcounter{page}{1}

\section{Introduction}

It is well known that compactification of higher-dimensional theories results in the vast landscape of vacua\cite{buosso, Kachru, susskind}. In the context of the multiverse, tunnelings between vacua become a relevant subject to study in order to answer the ultimate questions: why are we here, and where were we from?

While the real purpose is to bring string theory down to the low energy level, flux compactification in much simpler toy models has also been intensively studied\cite{gellman, partha, omero, salam, rubin}. In this simplified environment, one can make useful predictions while at the same time avoiding the mathematical complication. Recently, tunneling in the landscape of Einstein--Maxwell flux compactification was investigated~\cite{jjbp, carroll}. Their studies reveal novel channels of transdimensional transitions, both topology preserving and topology change. It is somewhat surprising that such a simple toy model can exhibit a rich landscape of vacua. 

It should be noted that in the context of realistic string compactification, the Einstein--Maxwell (and Einstein--Higgs) flux vacua provide a low-energy effective pictures. It is thus expected that at a high energy regime close to the string level, nonlinear terms should be taken into account. A natural nonlinear modification of the Maxwell Lagrangian is the Born-Infeld  theory~\cite{borninfeld}\footnote{Recently, the Born--Infeld-like modification has been employed in the study of noncanonical defects; for example, see Refs.~\cite{sarangi, babichev, pavlovski, ramadhan}.}; once constructed to regularize the self-energy divergence in the Maxwell theory, it has regained interest as the D-brane Lagrangian in the last 30 years. There is an extensive discussions on Einstein-Born-Infeld theories and their solutions in the literature. The Reissner--Nordstrom--Born--Infeld black hole is discussed in Refs~\cite{GSP, Breton, Fernando:2003tz, Fernando:2013uza}. More recently, this Einstein--Born--Infeld theory is generalized into Born--Infeld gravity coupled to Born--Infeld electrodynamics to study its corresponding static solutions and cosmological phenomena~\cite{Jana:2015cha, Bambi:2015sla}. 

A recent paper on Dirac--Born--Infeld (DBI)-gravity flux vacua~\cite{dbishiraishi} discusses the flux landscape in the context of DBI-type gravity theories. There, since the equations of motion do not lead to canonical Einstein equations, the authors employ a method of a minimum potential introduced by Wetterich~\cite{wetterich} to obtain minima which correspond to the flux vacua. Here, we follow a different route. In this paper, we consider Einstein--Born--Infeld theories and study the corresponding spontaneous compactification that leads to flux vacua. These models are the natural UV completion of the Einstein--Maxwell theory and can be perceived as one step closer toward the realistic string flux compactification.

In the next section, we discuss flux compactification in the Einstein-scalar-DBI (Dirac--Born--Infeld) theory, both in five and higher dimensions. Next we consider the six-dimensional Einstein--Born--Infeld--Abelian theory and study its flux vacua. In each case we discuss the effect of compactification on four-dimensional observers by means of dimensional reduction. Finally we in the last section, we give some comments on our results and discuss possible further work.

\section{Einstein-scalar--DBI Flux Compactifications}

\subsection{Five-dimensional model}
We consider a simple model of a complex scalar field in five dimensions as follows
\begin{equation}
{\cal S} = \int d\tilde{x}^{5}\sqrt{-\tilde{g}}\left(\frac{\bar{M}^{3}}{2}\tilde{R} -\beta^{2}\left(\sqrt{1 + \frac{1}{\beta^{2}}\partial_{M}\bar{\phi}\partial^{M}\phi} - 1\right) -\frac{\lambda}{4}(\bar{\phi}\phi - \eta^{2})^2 - \tilde{\Lambda}\right)\label{bz1},
\end{equation}
where  $M, N = 0, 1, 2, 3, 5$, with $\bar{M}$ and $\tilde{\Lambda}$ the five-dimensional Planck mass and cosmological constant, respectively. The complex scalar field is assumed to be static,  $\phi=\phi\left(x^M\right)=\eta e^{i\theta\left(x^M\right)}$, where $\theta(x^M)$ is the field phase. This is a nonlinear modification of the flux compactification model induced by scalars\footnote{The model can be perceived as flux compactification induced by a general nonlinear $\sigma$-model albeit with a different type of kinetic term than in Refs.~\cite{partha, skyrmecomp}.}~\cite{gellman,jjbp}. 

Assuming that the scalar fields are in their ground state, $|\phi|=\eta$,
\begin{equation}
S = \int d\tilde{x}^{5}\sqrt{-\tilde{g}}\left(\frac{\bar{M}^{3}}{2}\tilde{R} -\beta^{2}\left(\sqrt{1 + \frac{\eta^{2}}{\beta^{2}}\partial_{M}\theta\partial^{M}\theta} - 1\right) - \tilde{\Lambda}\right)\label{bz2},
\end{equation}
we obtain the equation of motion for our five-dimensional model
\begin{equation}
\tilde{R}_{AB} -\frac{1}{2}\tilde{g}_{AB}\tilde{R} = \frac{T_{AB}}{\bar{M}^{3}}\label{bz3},
\end{equation}
\begin{equation}
\partial_{M}\left(\frac{\sqrt{-\tilde{g}}\eta^{2}\partial_{M}\theta}{\sqrt{1 + \frac{\eta^{2}}{\beta^{2}}\partial_{A}\theta\partial^{A}\theta}}\right)= 0\label{bz4}.
\end{equation}

We employ the compactification ansatz,  
\begin{equation}
ds^{2} = \tilde{g}_{MN}dx^{M}dx^{N} = \tilde{g}_{\mu\nu}dx^{\mu}dx^{\nu} + \tilde{g}_{55}(x^{\mu})\left(dx^{5}\right)^2\label{bz5},
\end{equation}
with $\tilde{g}_{55}(x^{\mu}) = L^{2}=const.$, the size of the extra dimension. Here we take the four-dimensional Ricci scalar to be $R = 12H^{2}$, where $H$ is an effective four-dimensional Hubble constant; its value can be positive, negative, or zero for de Sitter, anti-de Sitter, and Minkowski, respectively~\cite{jjbp}. The Einstein tensor can then be written as
\begin{eqnarray}
\tilde{G}_{\mu\nu} &=& -3H^{2}\tilde{g}_{\mu\nu}\label{bz6}\\\tilde{G}_{55} &=& -6H^{2}\tilde{g}_{55}\label{bz7}.
\end{eqnarray}

The energy-momentum tensor is
\begin{equation}
T_{AB} = -\tilde{g}_{AB}\left(\beta^{2} \left(\sqrt{1 + \frac{\eta^{2}}{\beta^{2}}\partial_{M}\theta\partial^{M}\theta} - 1\right) + \tilde{\Lambda}\right) +\frac{\eta^{2}\partial_{A}\theta\partial_{B}\theta}{\sqrt{1 + \frac{\eta^{2}}{\beta^{2}}\partial_{M}\theta\partial^{M}\theta}}\label{bz8}.
\end{equation}
The scalar field ansatz is given by~\cite{gellman, jjbp}
\begin{equation}
\theta(x^{M})=nx^{5},
\end{equation} 
where $n$ is a winding number of which the value must be an integer by the definition of the field phase. In this sense, the scalar fields wrap around the extra dimension. This yields Einstein's equations in Eq. (\ref{bz3}) as follows:
\begin{eqnarray}
\tilde{G}_{\mu\nu} &=& -\frac{\tilde{g}_{\mu\nu}}{\bar{M}^{3}}\left(\beta^{2}\left(\sqrt{1 + \frac{\eta^{2}n^{2}}{\beta^{2}L^{2}}} - 1\right) + \tilde{\Lambda}\right)\nonumber\\
\tilde{G}_{55} &=& -\frac{\tilde{g}_{55}}{\bar{M}^{3}}\left(\beta^{2}\left(\sqrt{1 + \frac{\eta^{2}n^{2}}{\beta^{2}L^{2}}} - 1\right) + \tilde{\Lambda}\right) + \frac{\eta^{2}n^{2}}{\bar{M}^{3}\sqrt{1 + \frac{\eta^{2}n^{2}}{\beta^{2}L^{2}}}}\label{bz9}.
\end{eqnarray}

The solutions are
\begin{eqnarray}
(L^{2})_{\pm}&=&\frac{n^{2}\eta^{2}\left(3\beta^{4} +2\beta^{2}\tilde{\Lambda} -\tilde{\Lambda}^{2}\right) \pm n^{2}\eta^{2}\left(\beta^{2} -\tilde{\Lambda}\right)\sqrt{9\beta^{4} -2\beta^{2}\tilde{\Lambda} +\tilde{\Lambda}^{2}}}{2\left(\beta^{2}\tilde{\Lambda}^{2} -2\beta^{4}\tilde{\Lambda}\right)}\label{bz10}\\
(H^{2})_{\pm}&=&\frac{\tilde{\Lambda}+\frac{\beta^{2}}{4}\left(-4+\sqrt{2}\sqrt{\frac{5\beta^{4}+\tilde{\Lambda}\left(\tilde{\Lambda}\pm\sqrt{9\beta^{4}-2\beta^{2}\tilde{\Lambda}+\tilde{\Lambda}^{2}}\right)\mp\beta^{2}\left(\pm 2\tilde{\Lambda}+\sqrt{9\beta^{4}-2\beta^{2}\tilde{\Lambda}+\tilde{\Lambda}^{2}}\right)}{\beta^{4}}}\right)}{3\bar{M}^{3}}\label{bz11}.
\end{eqnarray}
For real values of $\eta$, $\beta$, $\tilde{\Lambda}$, and $\bar{M}$, it can be checked that:
\begin{itemize}
\item Only $(L^{2})_{+}$ gives real value, thus physical. 
\item We need negative $\tilde{\Lambda}$ to obtain real $H^{2}_{+}$. This means that compactification can occur provided that the five-dimensional bulk space is anti-de Sitter ($AdS_{5}$).
\end{itemize}

\subsection{Four-dimensional perspective}

The solutions above do not tell us about stability. It may be that they are saddle points of the energy and, thus, unstable. To understand stability we follow the approach of Refs.~\cite{jjbp, carroll}; that is, we look from the four-dimensional observer's perspective. We reduce the five-dimensional action (\ref{bz2}) into a four-dimensional effective theory using conformal metric
\begin{equation}
ds^{2} = \tilde{g}_{MN}dx^{M}dx^{N} = e^{\alpha\frac{\psi}{M_{p}}}g_{\mu\nu}dx^{\mu}dx^{\nu} + e^{\gamma\frac{\psi}{M_{p}}}L^{2}\left(dx^{5}\right)^2\label{by1}.
\end{equation}
The resulting lower-dimensional action takes the Einstein frame with canonical scalar radion terms should we choose $\alpha=-\sqrt{\frac{2}{3}}$ and $\gamma=2\sqrt{\frac{2}{3}}$; that is
\begin{equation}
S = \int d^{4}x\sqrt{-g}\left(\frac{M^{2}_{p}}{2}R - \frac{1}{2}\partial_{\mu}\psi\partial^{\mu}\psi - V(\psi,n)\right)\label{by2}
\end{equation}
with the Planck mass $M^{2}_{p}=2\pi L\bar{M}^{3}$. The last term is the radion ($\psi$) potential perceived from the point of view of the four-dimensional perspective,
\begin{equation}
V(\psi,n) = 2\pi L e^{-\sqrt{\frac{2}{3}}\frac{\psi}{M_{p}}}\left(\beta^{2}\left(\sqrt{1 + \frac{\eta^{2}}{\beta^{2}}\frac{n^{2}}{L^{2}}e^{-\sqrt{\frac{8}{3}}\frac{\psi}{M_{p}}}} - 1\right) + \tilde{\Lambda}\right)\label{by3}.
\end{equation}
The real radius of the compact extra dimension can be obtained by minimizing the potential,
\begin{equation}
\frac{dV(\psi)}{d\psi}\vert_{\psi=0}=0.
\end{equation}
The result confirms Eq. (\ref{bz11}).
\begin{figure}[htbp]
 \centering\leavevmode
 \epsfysize=6cm \epsfbox{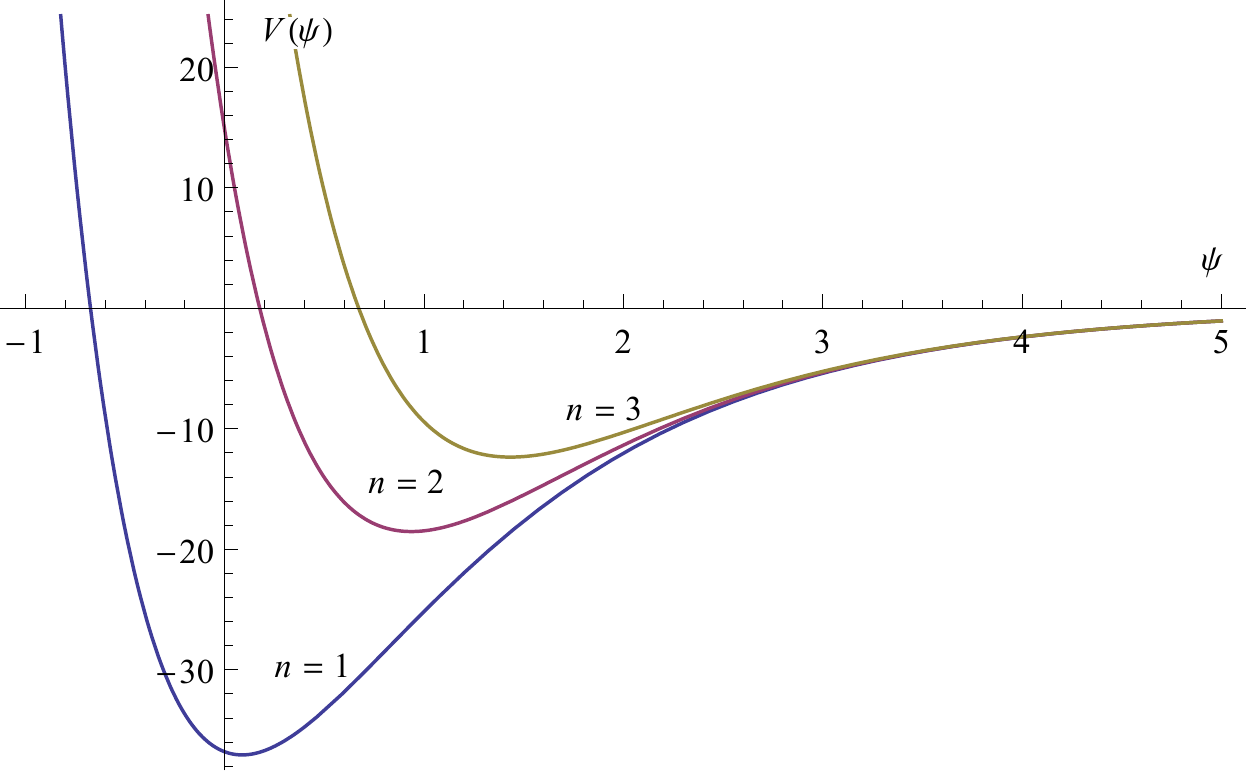}
 \caption {Plot of $V(\psi)$ vs $\psi$ for three different values of the winding number, $n=1,2,3,$ and other parameters as $M_{P}=\beta=1$, $\tilde{\Lambda}=-1$, and $\eta=L=10$.}
 \label{Vagnn}
\end{figure}
In Fig.~\ref{Vagnn} we plot the effective potential for various $n$. Upon Taylor expansion the first term in Eq. (\ref{by3}) apparently cannot produce higher-order coefficients that can stabilize against the repulsive force, in contrast to the model studied by Blanco-Pillado and Salem.~\cite{blancosalem}\footnote{In Ref.~\cite{blancosalem} the authors mentioned a DBI--Higgs model in (2+1+1) dimensions, which could be a more natural choice but nevertheless lacks any metastable (de Sitter) vacua. Apparently the same model in (3+1+1) dimensions also suffers from the absence of positive vacua.}. We are, therefore, forced to set the five-dimensional cosmological constant to be negative. Hence, the higher-dimensional space must be anti-de Sitter ($AdS_{5}$)

\begin{figure}
 \centering\leavevmode
 \epsfysize=8cm \epsfbox{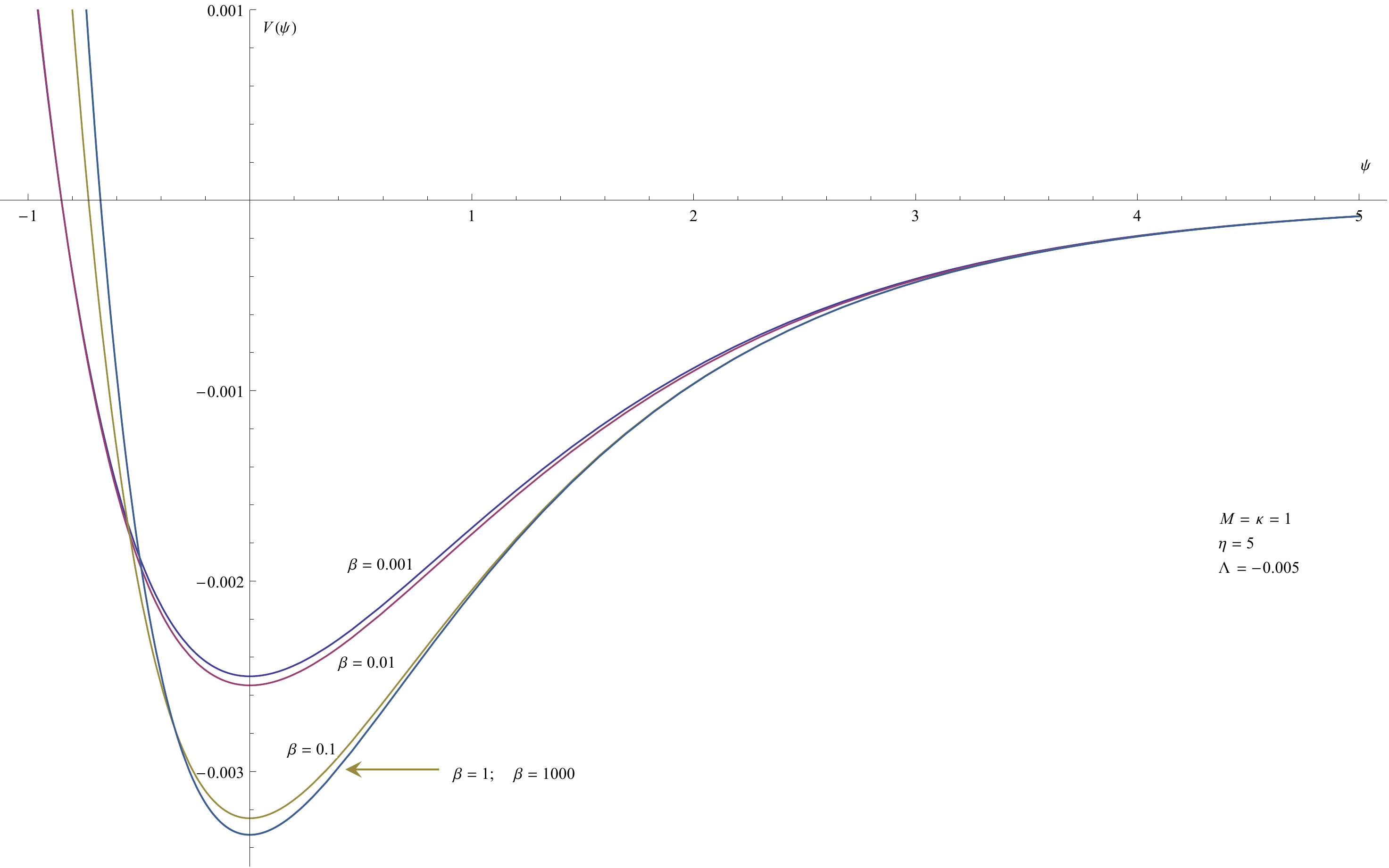}
 \caption {Plot of the effective potential $V(\psi)$ as a function of scalar field $\psi$ for four different values of the coupling constant. Here $M_{P}=1$, $n=1; $$\eta=5$, and $\Lambda=-0.005$.}
 \label{VagnBK}
\end{figure}

The distinctive feature this model has is its DBI nonlinearity controlled by the coupling constant $\beta$. As can be seen from Fig.~\ref{VagnBK}, the larger the $\beta$ (weaker coupling), the more stable the vacua are. This is simply because in the limit $\beta\rightarrow\infty$ our Lagrangian reduces to the ordinary five-dimensional Einstein--Higgs theory. On the other hand, for $\beta\rightarrow 0$ it enters the strong-coupling regime. In this regime the repulsive bosonic force increases, and hence the vacua tend to be less AdS.

\subsection{DBI-scalar in higher dimensions}
There is not much to say about the qualitative results of the potential shape. As in the case of the Einstein--Higgs theory~\cite{jjbp}, compactification only results in $AdS_4\times S^1$ vacua. But what happens if we have more than one extra dimension? Can we have $dS_4\times S^n$ vacua? In Ref. ~\cite{jjbp} the answer is negative. We are interested to see whether the nonlinear effect of Born--Infeld might cure this absence. Here we consider a $(4+q)$-dimensional theory where extra dimensions are compactified on a $q$-sphere,
\begin{equation}
S = \int d^{d}\tilde{x}\sqrt{-\tilde{g}}\left(\frac{\bar{M}^{d-2}}{2}\tilde{R}^{(d)} -\beta^{2}\left(\sqrt{1 + \frac{\eta^{2}}{\beta^{2}}k_{ij}\partial_{A}\phi^{i}\partial^{A}\bar{\phi}^{j}} - 1\right) - \tilde{\Lambda}\right),
\end{equation}
where $i,j=1,2,3,\dots,q$ label the extra dimensions and $k_{ij}(\phi^{k})$ is the field space metric on a $q$-sphere. We assume the conformal rescaling of the metric tensor
\begin{equation}
ds^{2} = e^{a\frac{\psi}{M_{P}}}g_{\mu\nu}dx^{\mu}dx^{\nu} + e^{b\frac{\psi}{M_{P}}}L^{2}h_{ij}dx^{i}dx^{j},
\end{equation}
where $L$ is the radius of the $q$-extra dimension and $a$ and $b$ are constants that depend on the extra dimension itself~\cite{jjbp},
\begin{align}
a& = -\sqrt{\frac{2q}{q+2}}& b& = 2\sqrt{\frac{2}{q(q+2)}}.&
\end{align}
The only scalar field ansatz consistent with the wrapping of the $q$-sphere is
\begin{equation}
\phi^{i}(\varphi^{i})=\varphi^{i}.
\end{equation}

The four-dimensional effective potential becomes
\begin{equation}
V(\psi,q) = -q(q-1)\frac{M^{2}_{P}}{2L^{2}}e^{-\frac{4}{qb}\frac{\psi}{M_{P}}} + V_{S}e^{\frac{a\psi}{M_{P}}}\left[\beta^{2}\left(\sqrt{1 +\frac{\eta^{2}q}{\beta^{2}L^{2}}e^{\frac{-b\psi}{M_{P}}}} -1\right) +\tilde{\Lambda}\right],\label{vpotq}
\end{equation}
where $V_{S}$ is the volume of a $q$-sphere with radius $L$. To obtain the radius of compactification, we solve the extremum condition of the potential, $\frac{dV}{d\psi}|_{\psi=0}=0$, which yields
\begin{equation}
\frac{q(q-1)}{L^{2}}\sqrt{\frac{q+2}{2q}} -\frac{\eta^{2}\kappa^{2}}{L^{2}\sqrt{1+\frac{q\eta^{2}}{\beta^{2}L^{2}}}}\sqrt{\frac{2q}{q+2}} -\kappa^{2}\sqrt{\frac{2q}{q+2}}\left(\beta^{2}\left(\sqrt{1+\frac{q\eta^{2}}{\beta^{2}L^{2}}}-1\right)+\tilde{\Lambda}\right)=0\label{z2}.
\end{equation}
This can be simplified as an algebraic equation of order $6$,
\begin{equation}
AL^{6}+BL^{4}+CL^{2}+D=0\label{z6},
\end{equation}
where
\begin{eqnarray}
A&\equiv& 4\kappa^{4}\beta^{2}\tilde{\Lambda}\left(2\beta^{2}-\tilde{\Lambda}\right),\nonumber\\
B&\equiv& 4\kappa^{2}(q+2)\left(\kappa^{2}\eta^{2}-(q-1)\right)\beta^{4} +4\kappa^{2}\tilde{\Lambda}\left(2q\kappa^{2}\eta^{2}+(q-1)(q+2)\right)\beta^{2} -4q\kappa^{4}\eta^{2}\tilde{\Lambda}^{2},\nonumber\\
C&\equiv& \left(4(q+1)^{2}\kappa^{4}\eta^{4}-(q-1)^{2}(q+2)^{2}-4q(q-1)(q+2)\kappa^{2}\eta^{2}\right)\beta^{2} +4q(q-1)(q+2)\eta^{2}\kappa^{2}\tilde{\Lambda},\nonumber\\
D&\equiv& -q\bigg((q-1)(q+2)\bigg)^{2}\eta^2.\label{z7}
\end{eqnarray}
There are six roots, three of which are real. The actual form is complicated and rather unilluminating, so we avoid presenting it. The root solution is the most positive one. We plug it back into Eq.~\eqref{vpotq} and study its behavior against the variation of $\beta$, $q$, and $\Lambda$.
\begin{figure}
 \centering
 \includegraphics[scale=0.5]{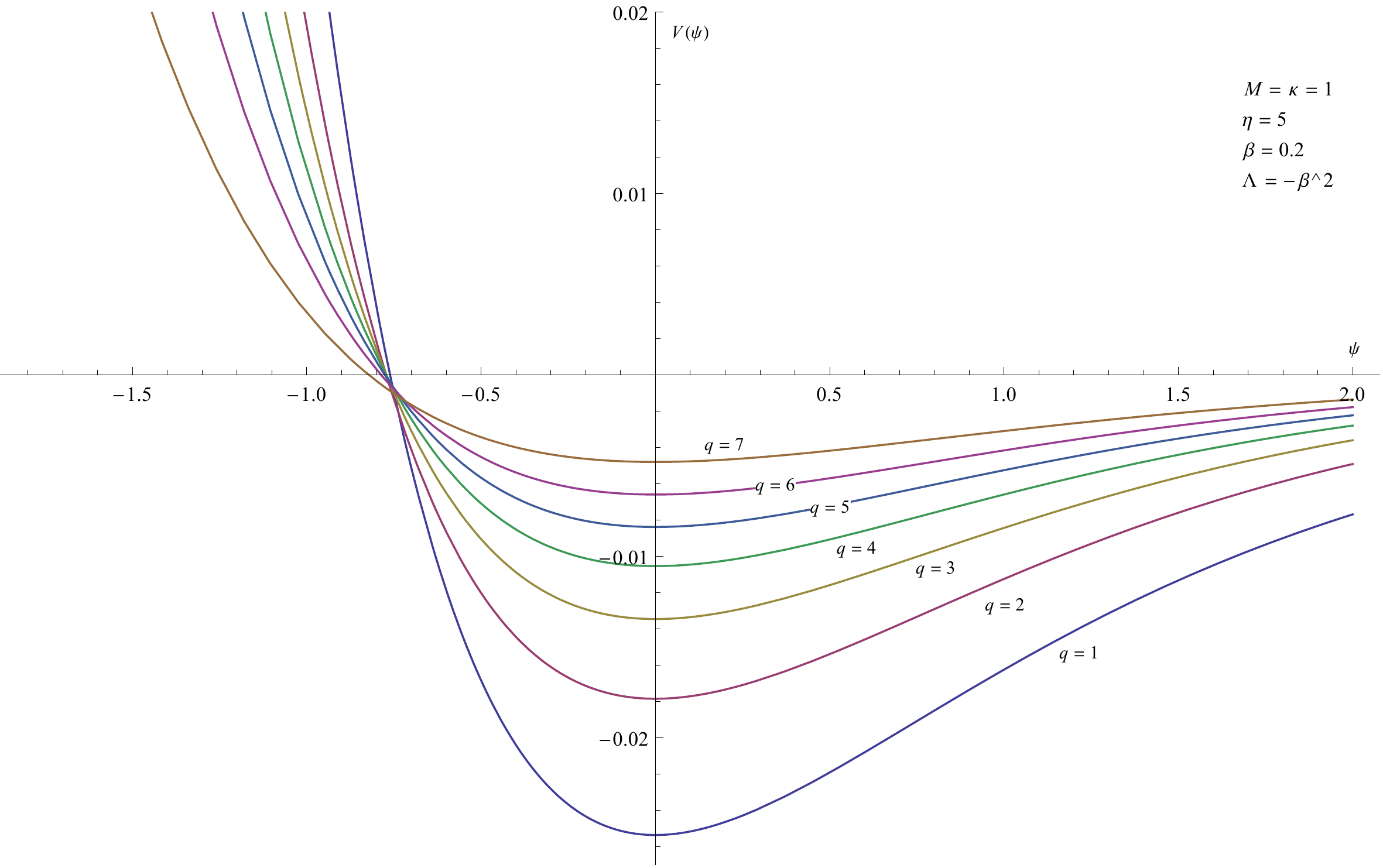}
 \caption{Plot of $V(\psi)$ from Eq.~\eqref{vpotq} with $M_P=\kappa=1$, $\eta=5$, $\beta=0.2$, and $\tilde{\Lambda}=-\beta^2$. In this parameter space compactification occurs only up to $q=7$.}
 \label{fig:Vlocex}
\end{figure}

It appears that the Born--Infeld nonlinearity still fails to produce any $\Lambda_{4d}\geq 0$-vacua. Compactification results in the $AdS_{4}\times S^{q}$ vacua. On the other hand, the existence of vacua now becomes a function of $\Lambda$ and $\beta$. This can be seen, for example, in Fig.~\ref{fig:Vlocex}. For every value of $\Lambda$ and $\beta$, there exists a critical number of the extra dimension, $q_{crit}$, beyond which no compactification solution is possible.

\begin{figure}
 \centering
 \includegraphics[scale=0.5]{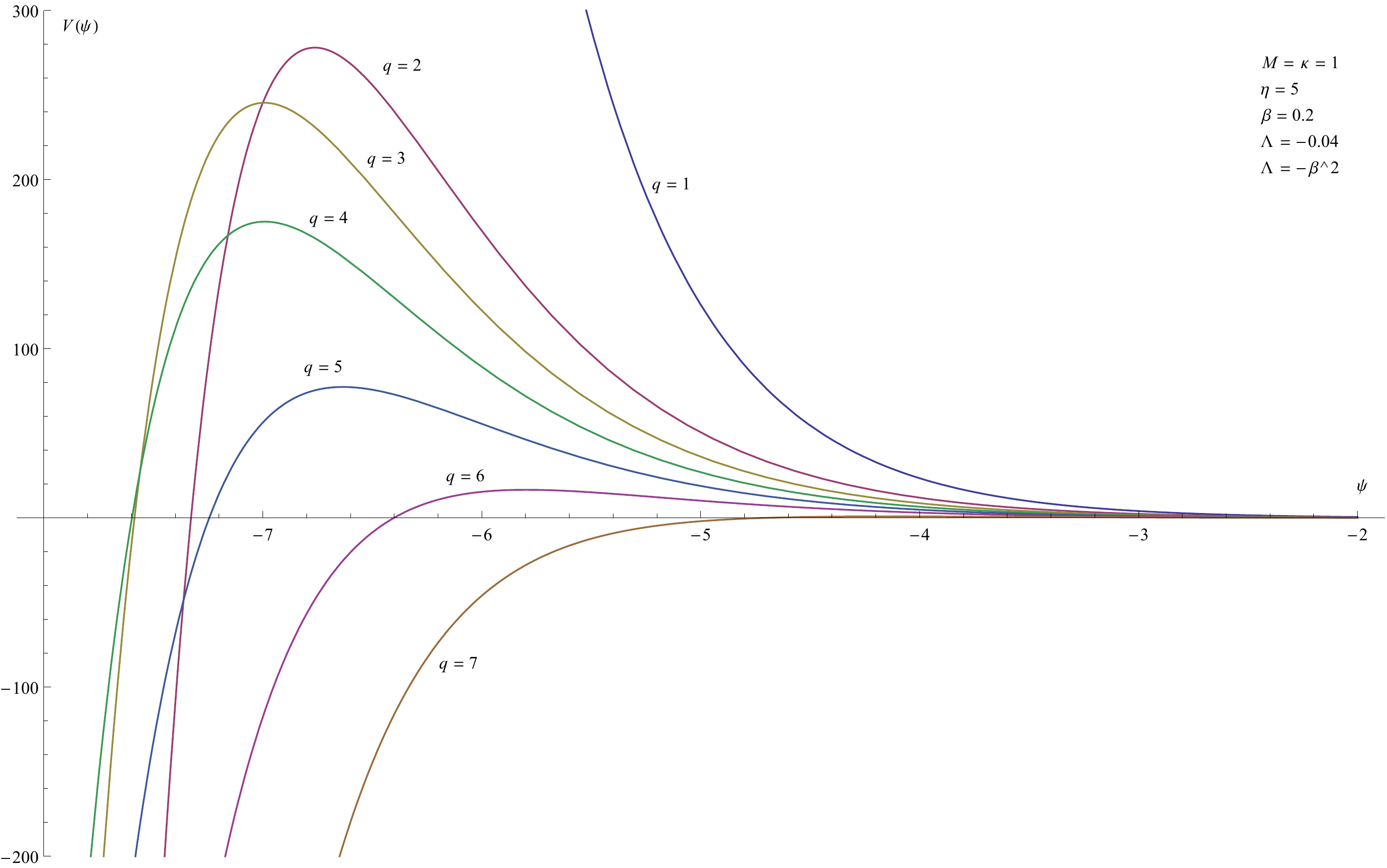}
 \caption{Global maxima of Fig.~\ref{fig:Vlocex}.}
 \label{fig:Vglobex}
\end{figure}

Another interesting property  is that, despite being perturbatively stable, all vacua with $q>1$ suffer from nonperturbative instability. This is because the vacua are all local minima of the potential (see Fig.~\ref{fig:Vglobex}). 
Spontaneously a bubble will nucleate, inside of which it contains a stabler vacuum~\cite{coleman}. But there is no stabler vacuum; the global minimum is at $V(\psi)\rightarrow-\infty$. In Ref.~\cite{browndahlen} this limit is identified as the limit where spacetime nonsingularly pinches off and nothing remains, a bubble of nothing\footnote{Recently bubbles of nothing in the flux compactification scenario have been discussed in Refs~\cite{bs, brs}.}~\cite{witten}. This instability is so severe that it renders all vacua to spontaneously tunnel to a state of no spacetime. Thus, we conclude that all compactification with $q>1$ extra dimensions is quantum-mechanically unstable against decay to nothing. This is a genuine feature of Einstein--Born--Infeld compactification that does not appear in the Einstein--Higgs theory.

Notice that the potential~\eqref{vpotq} contains an {\it effective} cosmological constant given by 
\begin{equation}
\Lambda_{eff}\equiv\tilde{\Lambda}-\beta^2.
\end{equation}
We investigate several regimes of $\Lambda_{eff}$ to see its effect on the compactification vacua:
\begin{figure}
 \centering
 \includegraphics[scale=0.5]{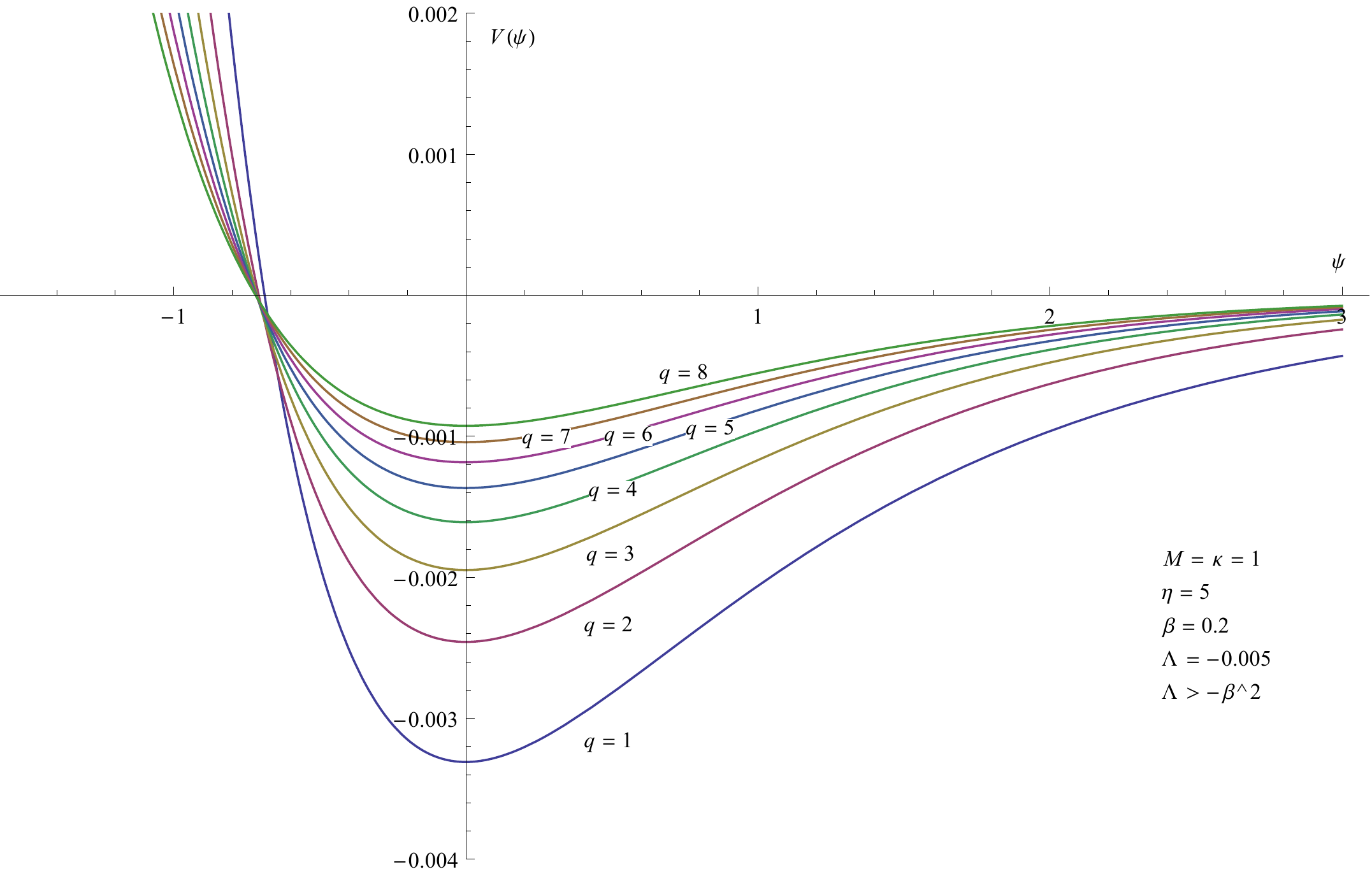}
 \caption{An example of $V(\psi)$ with $\Lambda_{eff}>-2\beta^2$. Here we use the following values: $M_P=\kappa=1$, $\eta=5$, $\beta=0.2$ and $\tilde{\Lambda}=-0.005$.}
 \label{fig:VLam>blex}
\end{figure}
\begin{enumerate}
\item $\Lambda_{eff}> -2\beta^2$\\
In this regime compactification can happen for a large class of $q$. This is shown, for example, in Fig.~\ref{fig:VLam>blex}.

\item $\Lambda_{eff}= -2\beta^2$\\
This is the regime where Fig.~\ref{fig:Vlocex} belongs. Here compactification can happen only up to $q=8$. From Fig.\ref{fig:Vglobex} it is shown that the most stable false vacua are when $q=2$.

\item $\Lambda_{eff}< -2\beta^2$\\
Here the strength of negative $\Lambda_{eff}$ results in fewer vacua possible. This is due to the attractive force from $AdS_{4}$. In this regime we can compactify solutions only up to $q=3$ (see Fig.~\ref{fig:VLam<blex}), 
\begin{figure}
 \centering
 \includegraphics[scale=0.5]{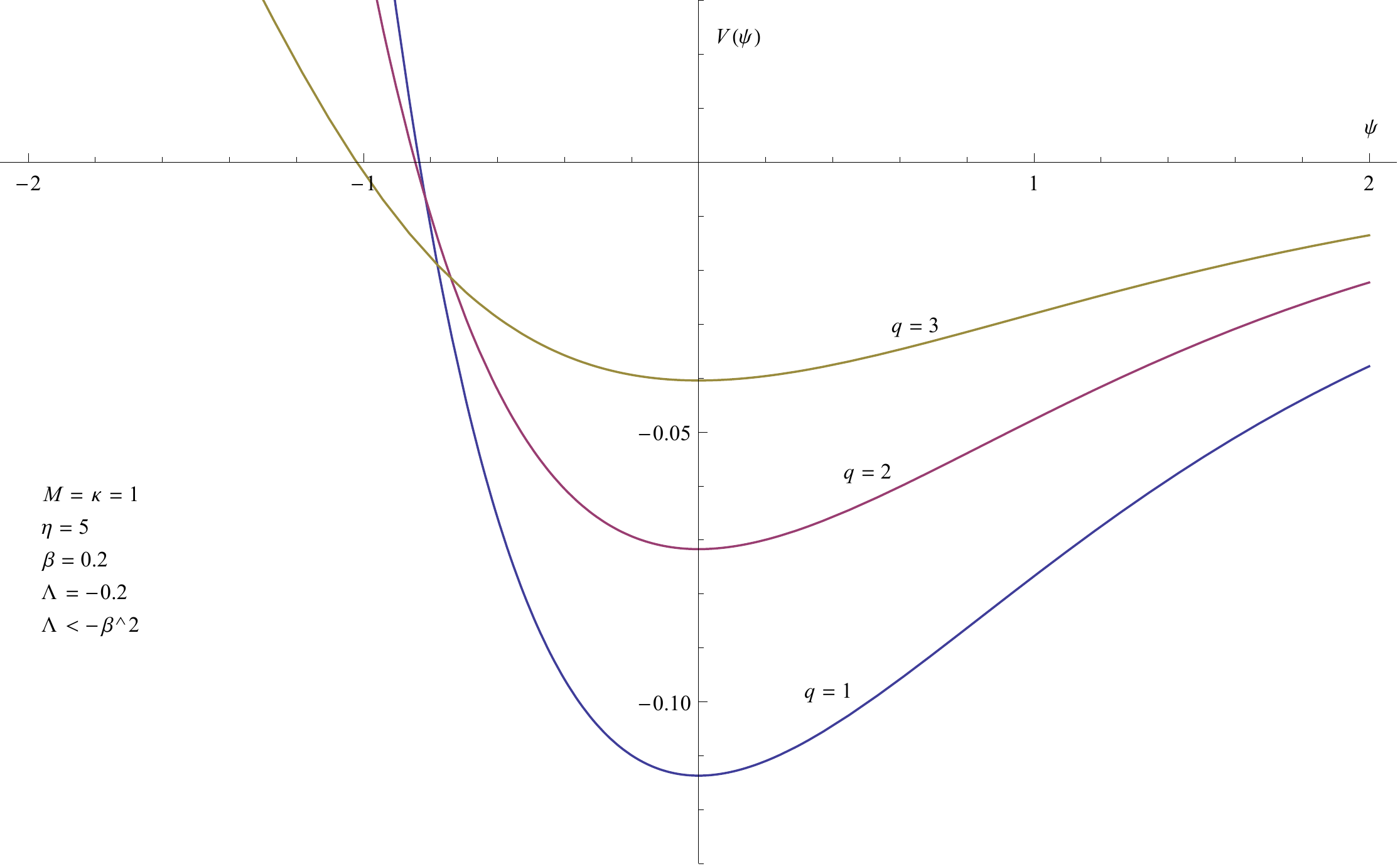}
 \caption{A typical $V(\psi)$ with $\Lambda_{eff}<-2\beta^2$. The following values are used: $M_P=\kappa=1$, $\eta=5$, $\beta=0.2$, and $\tilde{\Lambda}=-0.2$.}
 \label{fig:VLam<blex}
\end{figure}
beyond which no stable vacuum occurs. There even exists a critical limit of allowed extra dimensions above which the potential ceases to be real. We believe this is due to the square root form of the Born--Infeld term.
\end{enumerate}
\section{Landscape of Six-Dimensional Einstein-Abelian-DBI Theory}

In this section we consider a six-dimensional  generalization of Einstein-Maxwell theory~\cite{salam}. The action is given by the following:
\begin{equation}
\mathcal{S} = \int d^6 \tilde{x} \sqrt{-\tilde{g}} \left\{{\frac{M_{(6)}^4}{2} \tilde{R}^{(6)} - b^2 \left({\sqrt{1 + \frac{1}{2b^2} F_{MN}F^{MN}} + 1}\right) -\tilde{\Lambda}}\right\}, \label{zz}
\end{equation}
where  $M, N = 0, 1, 2, 3, 5, 6$ with $M_{(6)}$ and $\tilde{\Lambda}$ the six-dimensional Planck mass and cosmological constant, respectively. Varying it with respect to $g_{\mu\nu}$ and $A_{M}$ yields Einstein's field equations 
\begin{equation}
\label{eq:einsteineq}
\tilde{R}^{(6)}_{MN} - \frac{1}{2}\tilde{g}_{MN}\tilde{R}^{(6)} = \frac{1}{M_{(6)}^4} T_{MN}
\end{equation}
along with the energy-momentum tensor
\begin{equation}
T_{MN} = - \tilde{g}_{MN} b^2 \left({\sqrt{1 + \frac{1}{2b^2} F_{AB}F^{AB}} + 1}\right) + \frac{\tilde{g}^{CD}F_{MC} F_{ND}}{\sqrt{1 + \frac{1}{2b^2} F_{AB}F^{AB}}} + \tilde{g}_{MN} (b^2 - \tilde{\Lambda}),
\end{equation}
and Born--Infeld electrodynamics equation
\begin{equation}
\frac{1}{\sqrt{-\tilde{g}}}\partial_{N}\left(\frac{\sqrt{-\tilde{g}}F^{MN}}{\sqrt{1+\frac{1}{2b^2}F_{AB}F^{AB}}}\right)=0.
\end{equation}

The compactification ansatz for the metric is $X_4 \times S^2$,\footnote{$X_4$ can be $AdS_4$ (anti-de Sitter), $M_4$ (Minkowski), or $dS_4$ (de Sitter).}
\begin{equation}
ds^{2} = \tilde{g}_{MN}dx^{M}dx^{N} = \tilde{g}_{\mu\nu}dx^{\mu}dx^{\nu} + R^2\left(d\theta^2+\sin^2\theta d\varphi^2\right)\label{bz6},
\end{equation}
while the gauge field ansatz takes the form of the magnetic monopole~\cite{salam}
\begin{equation}
A_\varphi=-\frac{n}{2e}\left(\cos\theta\pm1\right).
\end{equation}
The Born--Infeld equation is trivially satisfied by the monopole ansatz, while Einstein's equations become
\begin{eqnarray}
-3 H^2 M^4_{(6)} - \frac{M^4_{(6)}}{R^2}&=&-b^2 \sqrt{1 + \frac{n^2}{4 b^2 e^2 R^4}} + b^2 - \tilde{\Lambda},\\ 
-6 H^2 M^4_{(6)} &=& - b^2 \sqrt{1 + \frac{n^2}{4 b^2 e^2 R^4}} + \frac{n^2}{4 e^2 R^4} \frac{1}{b^2 \sqrt{1 + \frac{n^2}{4 b^2 e^2 R^4}}} + b^2 - \tilde{\Lambda}. \label{2} 
\end{eqnarray}

Eliminating $H^{2}$, we obtain an algebraic equation for $R^{2}$ :

\begin{equation}
\left({\frac{2 M^4_{(6)}}{R^2} + b^2 - \tilde{\Lambda}}\right) \sqrt{1 + \frac{n^2}{4 b^2 e^2 R^4}} = b^2  + \frac{n^2}{2 e^2 R^4}.
\end{equation}
Rearranging, this gives a polynomial equation of order $4$ in $R^2$.
\begin{eqnarray}
\nonumber\frac{1}{R^8}\left({\frac{n^2 M^8_{(6)}}{b^2 e^2} - \frac{n^4}{4 e^4}}\right) + \frac{1}{R^6} \left({\frac{n^2 M^4_{(6)}(b^2 - \tilde{\Lambda})}{b^2 e^2}}\right) + \frac{1}{R^4} \left({4 M^8_{(6)} + \frac{n^2 (b^2 - \tilde{\Lambda})}{4 b^2 e^2} - \frac{b^2 n^2}{e^2}}\right) \\
+ \frac{1}{R^2} \left({4 M^4_{(6)} (b^2 - \tilde{\Lambda})}\right)+ (b^2 - \tilde{\Lambda})^2 - b^4 = 0,\label{a}
\end{eqnarray}
or, after some manipulation,
\begin{eqnarray}
Ax^4 + B x^3 + C x^2 + D x + F= 0,
\end{eqnarray}
where $x\equiv R^2$ and
\begin{eqnarray}
A&\equiv&\nonumber\left({(b^2 - \tilde{\Lambda})^2 - b^4}\right),\nonumber\\
B&\equiv&\left({4 M^4_{(6)} (b^2 - \tilde{\Lambda})}\right),\nonumber\\
C&\equiv&\left({4 M^8_{(6)} + \frac{n^2 (b^2 - \tilde{\Lambda})}{4 b^2 e^2} - \frac{b^2 n^2}{e^2}}\right),\nonumber\\
D&\equiv&\left({\frac{n^2 M^4_{(6)}(b^2 - \tilde{\Lambda})}{b^2 e^2}}\right),\nonumber\\
F&\equiv&\left(\frac{n^2 M^8_{(6)}}{b^2 e^2} - \frac{n^4}{4 e^4}\right).
\end{eqnarray}
There are four solutions of $R^2$, two of which are real. Their explicit forms are complicated, but it is clear that the true solution must be real and positive. Having solved the radius we can substitute back to obtain solutions for $H^2$. Whether the solutions describe the compactification or not can be best perceived by looking from the four-dimensional observer's point of view.

\subsection{Four-dimensional Perspective}

Like the previous way, for understanding the compactification mechanism and checking the stability, we shall look at this theory from a four-dimensional perspective where the radius of extra dimensions becomes a dynamical field (radion). We start by assuming that the form of the six-dimensional metric can be conformally rescaled as 
\begin{equation}
ds^2 = \tilde{g}_{MN} dx^M dx^N = e^{-\psi (x)/M_p} g_{\mu\nu} dx^\mu dx^\nu + e^{\psi (x)/M_p} R^2 d\Omega^2
\end{equation}

\begin{figure}
 \centering 
 \includegraphics[width=13cm,height=8cm]{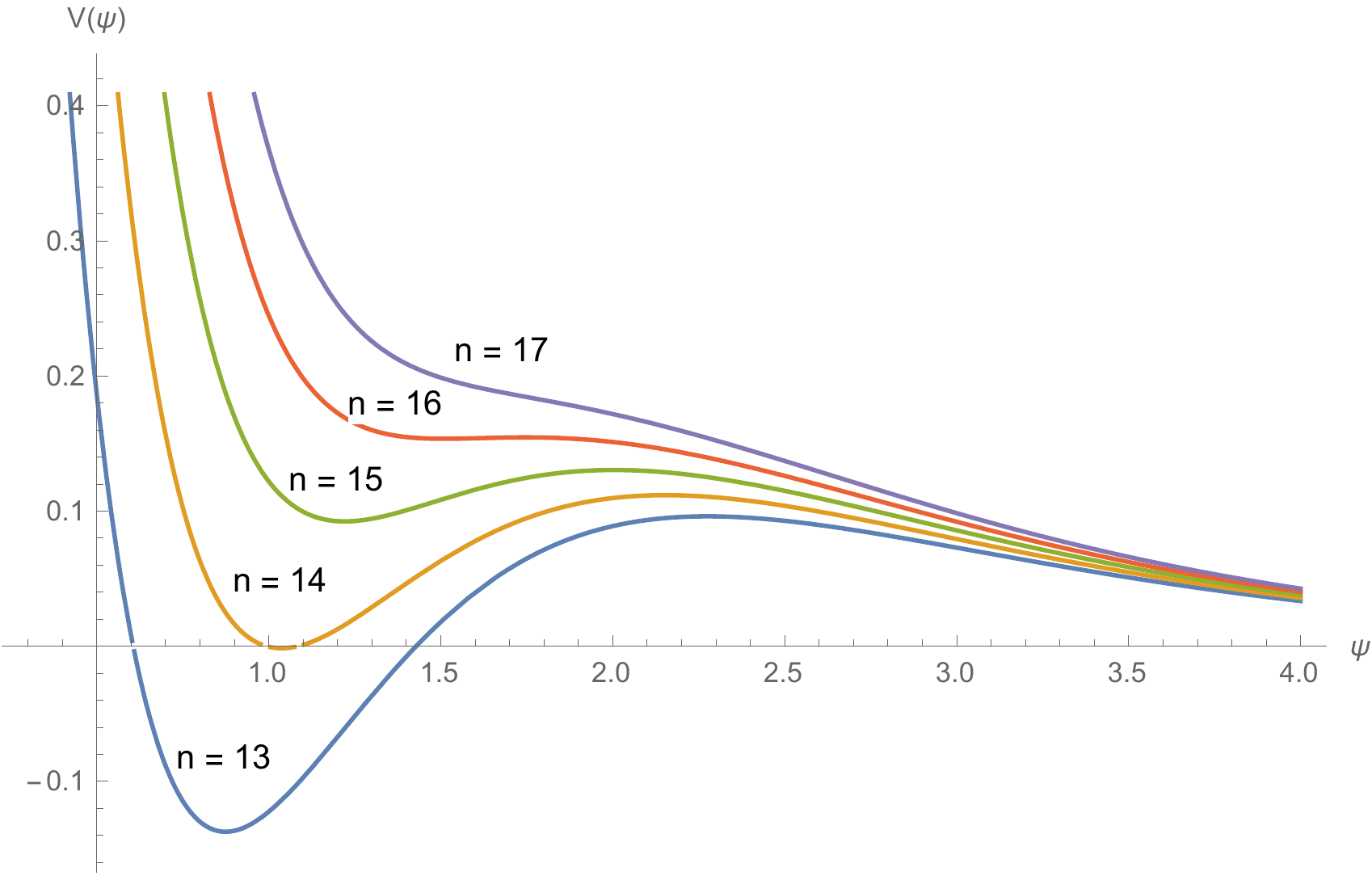}
 \caption{$\tilde{\Lambda} = 0.0104,~ b = 0.5,~n=13,14,15,16,17$}
 \label{thorough_case}
\end{figure}

After dimensional reduction the six-dimensional action can be written in a four-dimensional way as
\begin{eqnarray}
\mathcal{S} = \int d^4 x \sqrt{-g} \left\{{\frac{M_p^2}{2}R^{(4)} - \frac{1}{2} \partial_\mu\psi\partial^\mu\psi - V (\psi)}\right\}
\end{eqnarray}
with $V (\psi)$ given by
\begin{equation}
V(\psi)=4\pi M^4_{(6)} \left({\frac{b^2R^2}{M^4_{(6)}} \sqrt{1 + \frac{n^2 e^{-2\psi/M_p}}{4 b^2 e^2 R^4}} e^{-\psi/M_p} - e^{-2\psi/M_p} - \frac{(b^2 -\tilde{\Lambda})R^2}{M^4_{(6)}}e^{-\psi/M_p}}\right).\label{b}
\end{equation}
The first term comes from the Born--Infeld flux, that contributes to expanding the size of the extra dimensions. The second term comes from the curvature of $S^2$, while the last term is the constant term. Extremizing it yields Eq. (\ref{a}). Setting $M_6=M_p=e=1$, we can obtain the plot of $V(\psi)$ as shown in Fig. \ref{thorough_case}. We see from the picture that there are stable vacua which are stable against small perturbation and unstable ones.

These stable vacua, unlike the previous Einstein--DBI-Scalar case, can be positive, zero, or negative four-dimensional cosmological constant. The higher the winding number $n$ the greater the repulsive force due to the flux, thus the higher de Sitter vacua we have. The higher the de Sitter vacua the less stable they are against tunneling to decompactification ($dS_6$). There exists a critical value of $n$, $n_{crit}$, (in Fig.~\ref{thorough_case} $n_{crit}=17$),above which no compactification occurs.

\begin{figure}
 \centering 
 \includegraphics{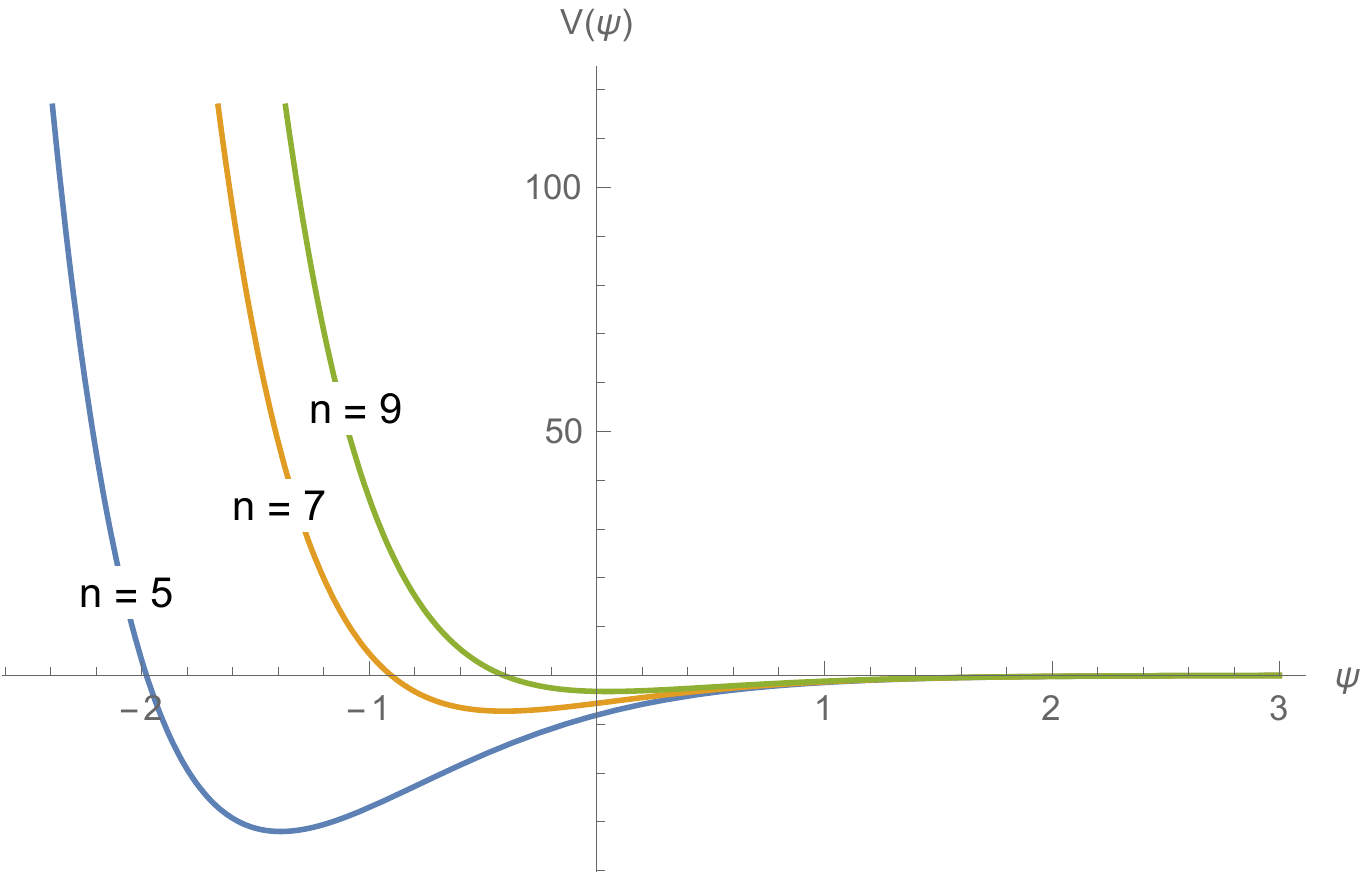}
 \caption{$\tilde{\Lambda}=0,~ b = 0.5,~n=5,7,10$. All lower-dimensional vacua are $AdS_4$.}
 \label{Lambda=0}
\end{figure}

Qualitatively, there is no nontrivial difference with the landscape of vacua shown in Ref. \cite{jjbp}. In this scenario, we are also forced to have $\tilde{\Lambda}$ to lift up vacua to obtain $dS_4$. Nonlinearity of the Born--Infeld kinetic term is not sufficient to self-lift up the vacua. As a consequence, all solutions with $\tilde{\Lambda} = 0$ result in $AdS_4$. This is shown, for example, in Fig. \ref{Lambda=0}.



\section{Conclusions}

In this paper we discussed flux vacua in the framework of Einstein--Born--Infeld theories. Our work was partly motivated by the search of nontrivial flux vacua in these theories, following the discovery of a surprisingly rich landscape in a simple higher-dimensional Einstein--Maxwell toy model~\cite{jjbp}; in particular we were curious whether the nonlinearity of Born--Infeld theories can avoid the need for a bulk cosmological constant and produce de Sitter vacua in the Einstein--Higgs toy model. It was expected that this UV completion of low-energy flux compactification might reveal some interesting features not shown in the linear (scalar and Maxwell) landscape. Our investigation results in the negative answer. The shapes of the four-dimensional radion potentials do not significantly differ from their linear counterparts, no de Sitter vacua appear in the Einstein--DBI--Higgs landscape irrespective of the bulk cosmological constant, and similarly so for the case of the Einstein--Abelian--DBI landscape. 

Despite having qualitatively same results, our investigation is not entirely barren. Perhaps the most nontrivial result is that we found all higher-dimensional scalar flux compactifications (with extra dimensions $q>1$) to be unstable against tunneling to, which we conjecture, nothing. This claim, however, requires a proof. Tunneling to nothing in the context of flux compactification has been extensively studied in Refs. \cite{bs, brs}. It is interesting to explicitly obtain instantons mediating decay to nothing in our toy model. We expect to present the result in the forthcoming publication.

Apart from tunneling to nothing, it is also interesting to investigate other decay channels,  e.g., tunneling between flux vacua. In Ref. \cite{jjbp} the instantons mediating flux tunneling in the Einstein--Maxwell toy model are identified as extremal magnetically charged $2$-branes. Our instanton should thus be the nonlinear modification of them. More specifically, it should be the higher-dimensional version of the Born--Infeld--Reissner--Nordstrom black hole. The solutions of Einstein--Abelian--Born--Infeld in four dimensions were already discovered in Refs. \cite{GSP, Breton, Fernando:2003tz, Fernando:2013uza}. However, so far the Born--Infeld brane solutions have not been found analytically. Once these solutions are known the construction of our toy-model instantons can be built along this line.

\section{Acknowledgments}

We thank Jose Juan Blanco-Pillado, Imam Fachruddin, Terry Mart, Agus Salam, and Anto Sulaksono for useful discussions and comments on this early manuscript. This work was partially supported by the University of Indonesia's Research Cluster Grant on ``Non-perturbative phenomena in nuclear astrophysics and cosmology" No.~1709/H2.R12/HKP.05.00/2014 and 1862/UN.R12/HKP.05.00/2015.


\end{document}